\documentclass[aps,prb,preprintnumbers,amsmath,amssymb,superscriptaddress,twocolumn,10pt]{revtex4-2}
\usepackage{txfonts}
\usepackage{graphicx}
\usepackage{dcolumn}
\usepackage{bm}
\usepackage{array}
\usepackage[dvipsnames]{xcolor}
\usepackage[%
    colorlinks=true,
    pdfborder={0 0 0},
    linkcolor=blue
]{hyperref}  
\usepackage{chngpage}
\usepackage{appendix}
\usepackage{subfig}
\usepackage{booktabs}

\usepackage{caption}
\newcommand{\ket}[1]{\lvert#1\rangle}

\newcommand{\be}{\begin{equation}}
\newcommand{\ee}{\end{equation}}
\newcommand{\bea}{\begin{eqnarray}}
\newcommand{\eea}{\end{eqnarray}}

\begin{document}
\title{Contrasting twisted bilayer graphene and  transition metal dichalcogenides for fractional Chern insulators: an emergent gauge picture}
\author{Heqiu Li}
\affiliation{%
Department of Physics, University of Toronto, Toronto, Ontario, Canada M5S 1A7
}
\author{Ying Su}
\affiliation{%
Theoretical Division, T-4 and CNLS, Los Alamos National Laboratory, Los Alamos, New Mexico 87545, USA
}
\affiliation{%
Center for Integrated Nanotechnologies (CINT), Los Alamos National Laboratory, Los Alamos, New Mexico 87545, USA
}
\author{Yong Baek Kim}
\affiliation{%
Department of Physics, University of Toronto, Toronto, Ontario, Canada M5S 1A7
}
\author{Hae-Young Kee}
\affiliation{%
Department of Physics, University of Toronto, Toronto, Ontario, Canada M5S 1A7
}
\affiliation{Canadian Institute for Advanced Research, CIFAR Program in Quantum Materials, Toronto, Ontario M5G 1M1, Canada}

\author{Kai Sun}
\email{sunkai@umich.edu}
\affiliation{%
Department of Physics, University of Michigan, Ann Arbor, MI 48109, USA
}

\author{Shi-Zeng Lin}
\email{szl@lanl.gov}
\affiliation{%
Theoretical Division, T-4 and CNLS, Los Alamos National Laboratory, Los Alamos, New Mexico 87545, USA
}
\affiliation{%
Center for Integrated Nanotechnologies (CINT), Los Alamos National Laboratory, Los Alamos, New Mexico 87545, USA
}

\begin{abstract}
The recent experimental discovery of the zero-field fractional Chern insulator (FCI) in twisted $\mathrm{MoTe_2}$ moir\'e superlattices has sparked immense interest in this exotic topological quantum state. The FCI has also been observed in previous experiments in magic angle twisted bilayer graphene (TBG) under a finite magnetic field of about 5 Tesla. Generally, the stabilization of FCI requires fine-tuning the topological band to satisfy certain conditions. It would still be helpful to have an intuitive picture to understand the different behaviors in twisted $\mathrm{MoTe_2}$ and TBG. Here, we compare them through the lens of emergent gauge fields. In TBG, the system can be mapped to two Dirac fermions coupled to emergent gauge fields with opposite signs. In contrast, the twisted $\mathrm{MoTe_2}$ reduces to a hole with parabolic dispersion coupled to an emergent gauge field. This contrasting gauge structure provides a new perspective on the observed difference: the zero-field FCI is stable in $\mathrm{MoTe_2}$ but absent in TBG. Based on this understanding, we will explore potential strategies for stabilizing FCI in both moir\'e superlattices.

\end{abstract}
\date{\today}
\maketitle

\section{Introduction}

Bands with nontrivial topology are exciting platforms for many exotic quantum states of matter. Examples include Chern insulators and topological insulators, where the bulk is fully gapped, but there exist topologically protected edge modes that are responsible for the quantized electrical Hall conductance. When the dispersion of the topological bands becomes narrow, in the so-called flat band limit in which the electron interaction is much larger than the bandwidth, new quantum states with topological order emerge at a fractional filling. This situation is similar to the fractional quantum Hall effect in Landau levels with Coulomb repulsion. The band version of the fractional quantum Hall effect is called fractional Chern insulators, which have been theoretically studied extensively in the last decade~\cite{PhysRevLett.106.236802,PhysRevLett.106.236803,PhysRevLett.106.236804,Regnault2011,Sheng2011,parameswaran_fractional_2013,bergholtz_topological_2013,PhysRevB.85.075116}. The fractional Chern insulators support anyons that carry fractional units of electric charge and unconventional quantum statistics and are proposed to have important applications in robust quantum computations.

Unlike the fractional quantum Hall effect, no external magnetic field is required to realize the FCI. Previous experiments report the existence of FCI in graphene on hexagonal boron nitride (hBN) under a strong magnetic field~\cite{Spanton_Zibrov_Zhou_Taniguchi_Watanabe_Zaletel_Young_2018}. The advent of moir\'e superlattice, which is achieved by misaligning two-dimensional materials, has offered an exciting new platform for exploring the FCI physics. The bands in the moir\'e superlattice can be tuned to be flat \cite{Bistritzer_MacDonald_2011} and topological \cite{PhysRevB.99.075127,PhysRevLett.122.086402} through gating, twist angle, etc.  The integer Chern insulators without magnetic field have been experimentally observed in the moir\'e superlattice, including magic-angle twisted bilayer graphene~\cite{Lu2019o,Serlin2020q}, and $\mathrm{MoTe_2}$ on $\mathrm{WSe_2}$~\cite{Li2021qa}. The signature of the FCI was reported in magic-angle twisted bilayer graphene for a magnetic field above 5 T~\cite{Xie_Pierce2021}. However, the FCI disappears at zero magnetic field. The breakthrough in observing zero-field FCI has been achieved very recently in twisted $\mathrm{MoTe_2}$ both in local spectroscopy~\cite{Cai2023f}, thermodynamic~\cite{Zeng2023e}, and later in transport measurements~\cite{Park2023f,Xu2023i}. The FCI in $\mathrm{MoTe_2}$ and other transition metal dichalcogenides (TMD) was first predicted in theory~\cite{Li2021fci,PhysRevB.107.L201109}, although at a slightly different angle, because of uncertainty in the material parameters. Later refined model parameters obtained from DFT calculations reproduce the FCI at a twisted angle consistent with experiments~\cite{Wang2023c}. Almost at the same time, the FCI has also been observed in pentalayer graphene on top of the hBN substrate~\cite{Lu_Han_Ju_2023}. These experimental breakthroughs have triggered tremendous excitement in the field, and many interesting theoretical ideas have emerged awaiting future experimental verification~\cite{PhysRevLett.131.136502,PhysRevLett.131.136501,Myerson-Jain_Jian_Xu_2023,Hu_Xiao_Ran_2023,Stern_Fu_2023,Song_Senthil_2023,Song_Jian_Fu_Xu_2023,Mao_Xu_Li_Bao_Liu_Xu_Felser_Fu_Zhang_2023,Yu_Herzog-Arbeitman_Wang_Vafek_Bernevig_Regnault_2023,Morales-Durán_Wei_Shi_MacDonald_2023,Jia_Yu_Liu_Herzog-Arbeitman_Qi_Regnault_Weng_Bernevig_Wu_2023,Herzog-Arbeitman_Wang_Liu_Tam_Qi_Jia_Efetov_Vafek_2023,Kwan_Yu_Herzog-Arbeitman_Efetov_Regnault_Bernevig_2023,Dong_Patri_Senthil_2023,Zhou_Yang_Zhang_2023,Guo_Lu_Xie_Liu_2023}.

The conditions for FCI in the Chern band have been well established by taking an analogy to the Landau levels, which are summarized as follows \cite{PhysRevB.86.165129,PhysRevB.90.165139,parameswaran_fractional_2013,PhysRevResearch.2.023237}. 1) The topological band should be sufficiently flat compared to the Coulomb interaction energy scale. 2) The gap between the band of interest and the other bands should be larger than the Coulomb interaction. 3) The Berry curvature distribution should be sufficiently homogeneous in the Brillouin zone. 4) the so-called trace condition, where the trace of the quantum metric tensor should be equal to the Berry curvature for any momentum in the Brillouin zone. When one turns off the interlayer tunneling between the same sublattice in TBG, the so-called chiral limit~\cite{Tarnopolsky2019m}, the conditions 1), 2) and 4) are satisfied, and the flat band closely resembles the Landau levels, and FCI is expected in this limit. Indeed, previous numerical calculations predicted the zero-field FCI in magic-angle twisted bilayer graphene. When the ratio of the interlayer tunneling between the same sublattice and different sublattice becomes large, it is shown that the charge density wave (CDW) is stabilized instead~\cite{Wilhelm2021f}. The absence of the zero-field FCI in TBG in the experiment was attributed to the violation of the aforementioned conditions due to the large deviation from the chiral limit. However, a more intuitive picture is still lacking. We believe that such an intuitive understanding is important for searching for new material systems for FCI. With this in mind, we will fill this gap in this work. In particular, we will focus on the conditions under which the moir\'e superlattice can be mapped to electrons coupled to an emergent U(1) gauge field, which can make the comparison with the fractional quantum Hall effect more transparent.

When comparing TBG and TMD, one obvious difference is the spin-valley locking in TMD. One may wonder if the spin-orbit coupling (SOC) which is responsible for the spin-valley locking is the main factor for the appearance of the zero fields FCI in TMD. This is not the case as we will show in this work. We provide an intuitive picture of why FCI is not favored in TBG by invoking the emergent gauge approach. We compare the emergent gauge structure in TBG and TMD, through which it becomes transparent on the conditions for FCI in TMD and TBG. In TBG, on a proper basis, the Hamiltonian for a given valley is cast into two Dirac fermions coupled to an opposite emergent magnetic field. The two Dirac fermions are then coupled through the interlayer tunneling between the same sublattice. In  TMD, the Hamiltonian at a given valley is described by a double exchange model with classical localized layer pseudospins forming a skyrmion spin texture. In the strong coupling limit, the Hamiltonian becomes an electron orbitally coupled to an emergent U(1) gauge field. {This different emergent gauge structure sheds new light on the experimental fact that no zero-field FCI was observed in TBG, while it was observed in TMD, and points to different strategies to achieve FCI in these two systems.}

The remainder of the manuscript is organized as follows. In Sec. II, we introduce the continuum model for TBG, and unravel the hidden gauge structure. We will benchmark the flat topological band within this picture, and calculate the phase diagram when we turn on the interlayer tunneling between the same sublattice. We will then discuss the role of SOC in TBG, which can be introduced by coupling the TBG to the substrate with strong SOC, such as $\mathrm{WSe_2}$, and its impact on FCI. In Sec. III, we consider the continuum model for TMD in terms of a gauge field. The paper is concluded by discussion and summary in Sec. IV. 

\begin{figure}
\centering
\includegraphics[width=3.4 in]{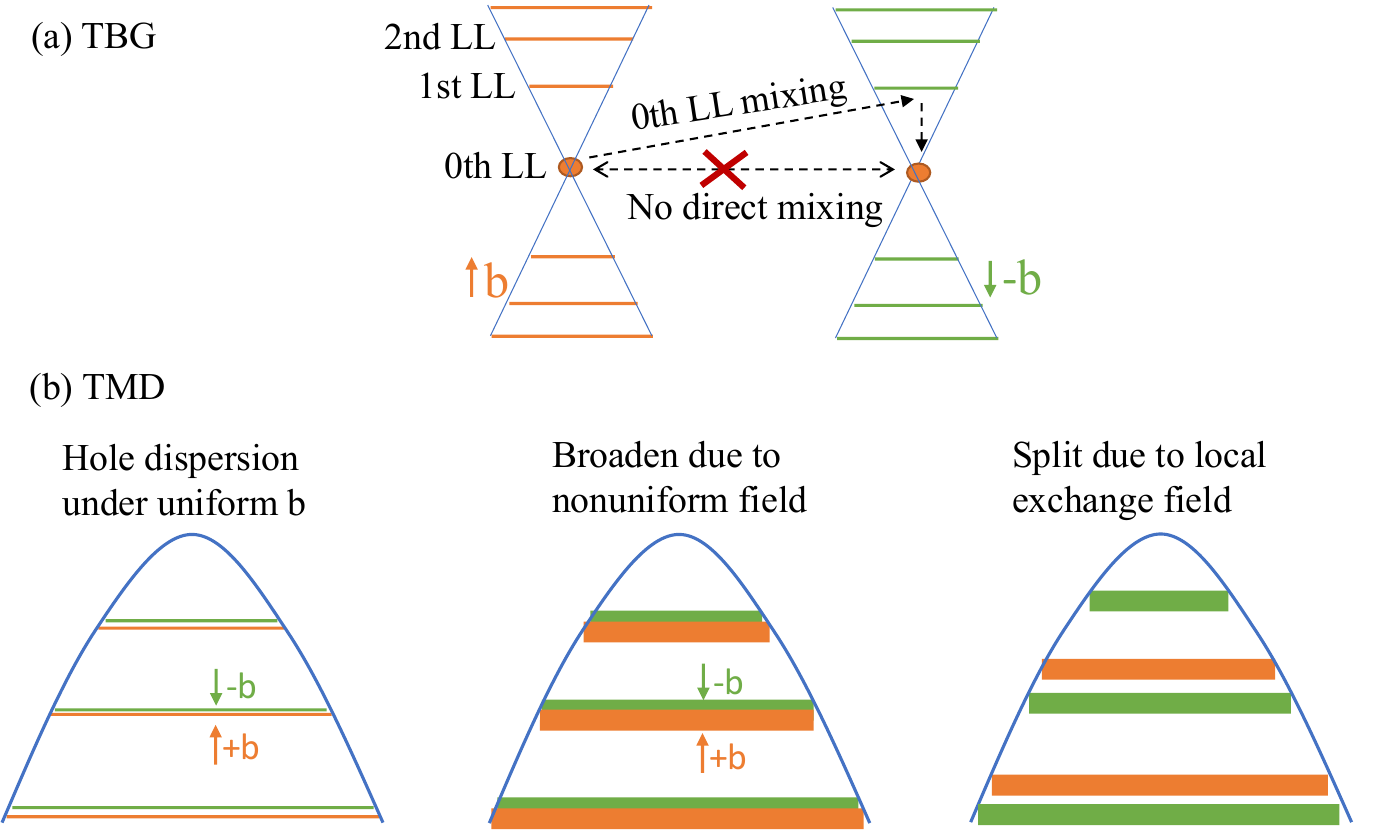}
\caption{(a) In TBG near the AA region, the valley resolved Hamiltonian describes two Dirac cones coupled to an opposite emergent magnetic field. The emergent field stabilizes the Landau levels and the zeroth Landau level is sublattice polarized. The zeroth Landau levels with opposite emergent fields hybridize through higher Landau levels via the $w_{AA}$ interlayer coupling. (b) In TMD moir\'e superlattice, the valley resolved Hamiltonian maps to two hole bands coupled to an emergent magnetic field with opposite sign due to the moir\'e potential. The nonuniform emergent field broadens the Landau levels. Furthermore, the nonuniform exchange coupling splits the Landau levels associated with the opposite magnetic field.}
\label{figTMD}
\end{figure}

\section{Twisted bilayer graphene}
Here we map the Hamiltonian of twisted bilayer graphene to the Dirac fermion coupled to an emergent gauge field. The continuum Hamiltonian for the twisted bilayer graphene (TBG) describes the hybridization of two Dirac fermions at $\kappa_t$ ($\kappa_b$) momentum in the top (bottom) layer. For the $\zeta=+$ valley, the Hamiltonian is
\begin{equation}\label{neq1}
    \mathcal{H}(\mathbf{k})_+=
    \begin{bmatrix}
-v_F(\mathbf{k}-\mathbf{\kappa}_t)\cdot\mathbf{\sigma}  & \Delta_T \exp(-i \mathbf{\Delta}_K\cdot \mathbf{r})\\
\Delta_T^\dagger \exp(i \mathbf{\Delta}_K\cdot \mathbf{r})&  -v_F(\mathbf{k}-\mathbf{\kappa}_b)\cdot\mathbf{\sigma} 
\end{bmatrix},
\end{equation}
where $v_F$ is the Fermi velocity, $\mathbf{k}=(k_x,\ k_y)$ is the momentum, and $\Delta_T \exp(-i \mathbf{\Delta}_K\cdot \mathbf{r})$ describes the interlayer tunneling with $\mathbf{\Delta}_k\equiv \mathbf{\kappa}_b - \mathbf{\kappa}_t$. The Pauli matrices $\mathbf{\sigma}=(-\sigma_x,\ \sigma_y)$ act on the (A, B) sublattice degree of freedom for graphene. The interlayer coupling is 
\begin{equation}
    \Delta_T(\mathbf{r})=
    \begin{bmatrix}
w_{AA} g(\mathbf{r})  &  w_{AB} g(\mathbf{r}-\mathbf{r}_{AB})\\
w_{AB} g(\mathbf{r}+\mathbf{r}_{AB}) &  w_{AA} g(\mathbf{r}) 
\end{bmatrix},
\end{equation}
with $g(\mathbf{r})=\sum_{j=1}^3 \exp(i \mathbf{q}_j\cdot \mathbf{r})$ and $\mathbf{q}_1=(0, 4\pi/3 L_s)$, $\mathbf{q}_2=(-2\pi/\sqrt{3} L_s, -2\pi/3L_s)$, $\mathbf{q}_3=(2\pi/\sqrt{3} L_s, -2\pi/3L_s)$. Here $L_s$ is the moir\'e supercell size, $w_{AA}$ and $w_{AB}$ are the tunneling strength between the AA and AB region, respectively. The model has $C_{3z}$, $C_{2x}$ and $C_{2z}\mathcal{T}$ symmetries, where $\mathcal{T}$ is the spinless time reversal operation. The Hamiltonian at the $\zeta=-$ valley is obtained by $\mathcal{T}$.

We can remove the momentum offset $\kappa_{t/b}$ in Eq. \eqref{neq1} by the following gauge transformation $\Psi_{t/b}\rightarrow \Psi_{t/b}\exp(i \mathbf{\kappa}_{t/b} \cdot \mathbf{r})$, where $\Psi_{t/b}$ are the single particle wave function in the top/bottom layer. In the new basis, the Hamiltonian becomes
\begin{equation}\label{eq3nn}
    \mathcal{H}(k)_+=
    \begin{bmatrix}
-v_F \mathbf{k}\cdot\mathbf{\sigma}  & \Delta_T\\
\Delta_T^\dagger &  -v_F \mathbf{k}\cdot\mathbf{\sigma} 
\end{bmatrix}.
\end{equation}
Equation \eqref{eq3nn} describes Dirac fermions coupled to an SU(2) gauge field, see Appendix \ref{AppenA} for a detailed discussion. Because the quantum states in the TBG flat bands are mostly localized near the AA stacking region~\cite{Cao_Fatemi_Demir_Fang_Tomarken2018}, we focus on the AA region. Remarkably, within the vicinity of the AA point, we find that the SU(2) gauge fields are parallel in the vector space spanned by the generators of the SU(2) gauge fields over an extended region $\mathbf{r}$. In this case, the SU(2) gauge fields can be diagonalized, and the model is reduced to two opposite U(1) gauge fields. Near the AA region when $r\ll L_s$, we can expand $\Delta_T(\mathbf{r})$  to the linear order in $\mathbf{r}$ \cite{PhysRevB.99.155415}
\begin{equation}\label{eq3}
    \mathcal{H}(\mathbf{k})_+=-v_F(\mathbf{k}-\mathbf{a} \tau_y)\cdot\mathbf{\sigma}+3 w_{AA}\tau_x,
\end{equation}
where the emergent gauge field is $\mathbf{a}=2\pi w_{AB}(y,\ -x)/L_sv_F$, and the Pauli matrices $\tau$ act on the layer pseudospin. At the magic angle $\theta=1.08^\circ$, the emergent magnetic field is $b=(\nabla\times \mathbf{a})_z\approx 120$ T. In the $\zeta=-$ valley,  the emergent magnetic field is opposite to that of the $\zeta=+$ valley. Rotating the layer pseudospin $\tau_y\rightarrow \tau_z$ and $\tau_x\rightarrow \tau_y$, the Hamiltonian becomes
\begin{equation}\label{eq4}
    \mathcal{H}(\mathbf{k})_+=
    \begin{bmatrix}
-v_F(\mathbf{k}-\mathbf{a})\cdot\mathbf{\sigma}  & -3i w_{AA}\\
3i w_{AA} &  -v_F (\mathbf{k}+\mathbf{a})\cdot\mathbf{\sigma} 
\end{bmatrix}.
\end{equation}
Now it becomes clear that the TBG near AA region can be regarded as two Dirac fermions coupled independently to the opposite gauge field. These two Dirac fermions are coupled through the interlayer coupling $w_{AA}$. 

In the chiral limit when $w_{AA}=0$, we have the zeroth Landau level due to the $\mathbf{a}$ and $-\mathbf{a}$ gauge fields, and they are degenerate and sublattice polarized at energy $E=0$. In the presence of Coulomb interaction, the system polarizes to a distinct Chern basis determined by the sign of $\mathbf{a}$. Away from the chiral limit when $w_{AA}\neq 0$, the wave functions of $\mathbf{a}$ and $-\mathbf{a}$ blocks start to hybridize and open a gap in the spectrum. A direct hybridization between the zeroth Landau level for the $\mathbf{a}$ and $-\mathbf{a}$ gauge fields is not allowed because they are polarized in the different sublattices. The hybridization happens through high order Landau levels, and is of the order of $(w_{AA}/\hbar \omega_c)^2$ where $\hbar \omega_c$ is the energy gap between the zeroth and first Landau level. The hybridization causes mixing of Landau levels with an opposite Chern number, thus disfavoring the FCI state.

\subsection{Numerical results}
\begin{figure}
\centering
\includegraphics[width=3.4 in]{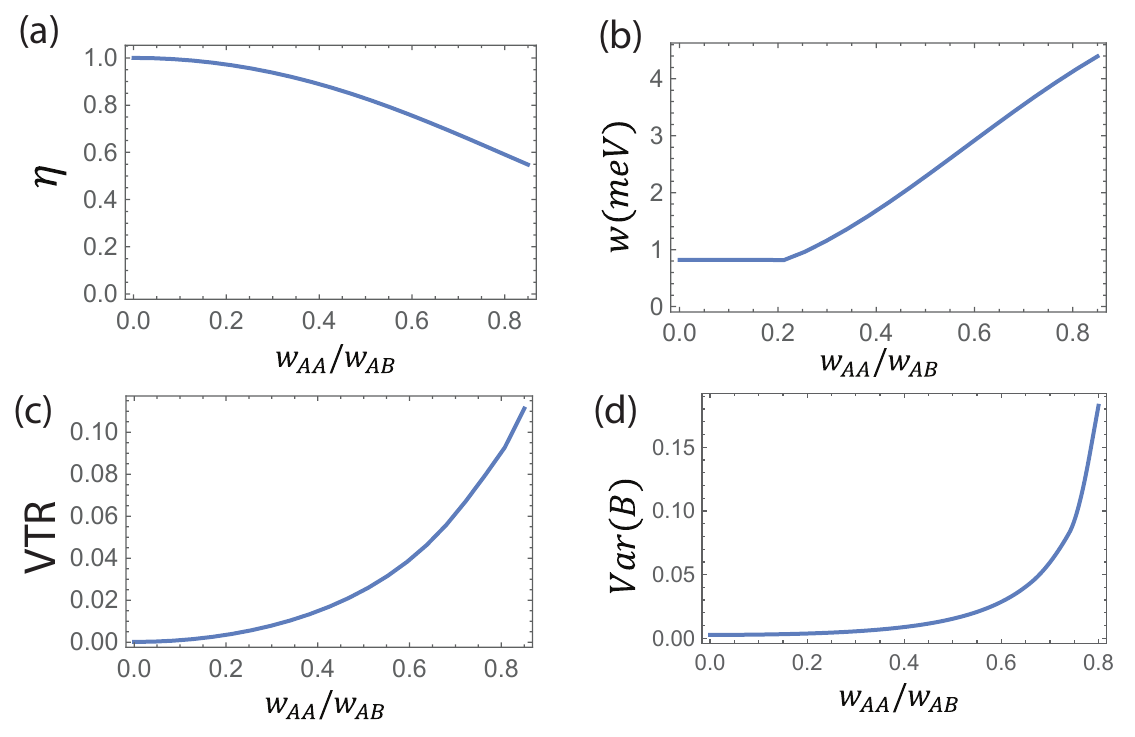}
\caption{(a) Overlap $\eta$ between eigenstates and the ideal zeroth Landau level for the Landau level model of TBG/hBN at different ratio $w_{AA}/w_{AB}$. (b) Bandwidth of the flat band in TBG/hBN at different chiral ratio $w_{AA}/w_{AB}$. (c) Violation of trace condition (VTR) and (d) variance of Berry curvature at different ratio $w_{AA}/w_{AB}$. The larger ratio $w_{AA}/w_{AB}$ is accompanied with larger bandwidth, larger violation of trace condition and variance of Berry curvature, hence FCI is disfavored.  }
\label{fig_chiral}
\end{figure}

We compute the mixing of the zeroth Landau levels with the opposite $\mathbf{a}$ by solving the Landau level problem in Eq. \eqref{eq4}. We quantify the Landau level mixing by considering the overlap $\eta$ between the low energy eigenstate at finite $w_{AA}$ and the exact zeroth Landau level at $w_{AA}=0$. The precise definition of $\eta$ is given in the Appendix \ref{AppenB}. {blue}{We also take into account the aligned hBN substrate, which provides a sublattice potential $m\sigma_z$ to one graphene layer with $m=15$ meV~\cite{PhysRevResearch.1.033126}.} The results for $\eta$ versus $w_{AA}/w_{AB}$ are shown in Fig. \ref{fig_chiral}, where $1-\eta\sim w_{AA}^2$ for a small $w_{AA}$. We also compute the bandwidth, trace condition variation defined as $\frac{1}{\cal A}\int_{\mathrm{BZ}}(\mathrm{tr}[g(\mathbf{k})]-|B(\mathbf{k})|)d^2k$ and Berry curvature variation $\frac{1}{\cal A}\int_{\mathrm{BZ}}\left(B(\mathbf{k})-\overline{B}\right)^2 d^2k$ directly from the continuum model of moir\'e superlattice, where $\cal A$ is the area of Brillouin zone. When the mixing between the two zeroth Landau levels increases by increasing $w_{AA}/w_{AB}$, the band width increases, the Berry curvature distribution becomes more nonuniform and the violation of the trace condition becomes stronger.

We also perform exact diagonalization (ED) calculations of TBG/hBN by projecting the Hamiltonian into the flat bands. To study the effect of $w_{AA}/w_{AB}$ on FCI, we focus on the valley and spin polarized sector. The interacting Hamiltonian is
\bea
H_\mathrm{I}=\frac{1}{2\Omega}\sum_{|\mathbf q+\mathbf G|\neq 0} V(\mathbf q+\mathbf G):\rho_{\mathbf q+\mathbf G}\rho_{-\mathbf q-\mathbf G}:,\ 
\label{eqHint}
\eea
where $\Omega$ is the system area, $V(\mathbf{q})=\frac{\mathrm{e}^{2}}{2 \varepsilon} \frac{1}{|\mathbf q|}\left(1-e^{-|\mathbf q|d_0}\right)$, $\rho_{\mathbf q}=\sum_{\mathbf k, m_{1}, m_{2},\zeta,s}\lambda_{m_1,m_2,\zeta,s}(\mathbf k,\mathbf k+\mathbf q) \left(d_{\mathbf k, m_{1},\zeta,s}^{\dagger} d_{\mathbf k+\mathbf q, m_{2},\zeta,s}  \right)$, and the form factor $\lambda_{m_{1}, m_{2},\zeta,s}(\mathbf{k}, \mathbf{k}+\mathbf{q}) \equiv\left\langle u_{\mathbf k, m_1,\zeta,s} \mid u_{\mathbf k+\mathbf q, m_2,\zeta,s}\right\rangle$. Here $:\ :$ stands for normal ordering, $m_1,m_2$ are band indices, and $\mathbf{G}$ is the moir\'e reciprocal lattice vector. We take the dielectric constant $\epsilon=7\epsilon_0$, $d_0=5L_s$ and the twist angle $\theta=1.08^\circ$ in our computation. To fully characterize the ground state, we compute the energy spectrum, the density-density correction $\chi(\mathbf{q})=\langle \rho(\mathbf{q})\rho(-\mathbf{q})\rangle-\langle \rho(\mathbf{q})\rangle\langle \rho(-\mathbf{q})\rangle$ and the particle entanglement spectrum (ES). To calculate ES, we divide the $N$ particles into $N_A$ and $N_B=N-N_A$ particles and trace out $N_B$ particles to get the reduced density matrix $\rho_A$. The ES levels $\xi$ are obtained from the logarithm of eigenvalues of $\rho_A$. 

\begin{figure}
\centering
\includegraphics[width=3.4 in]{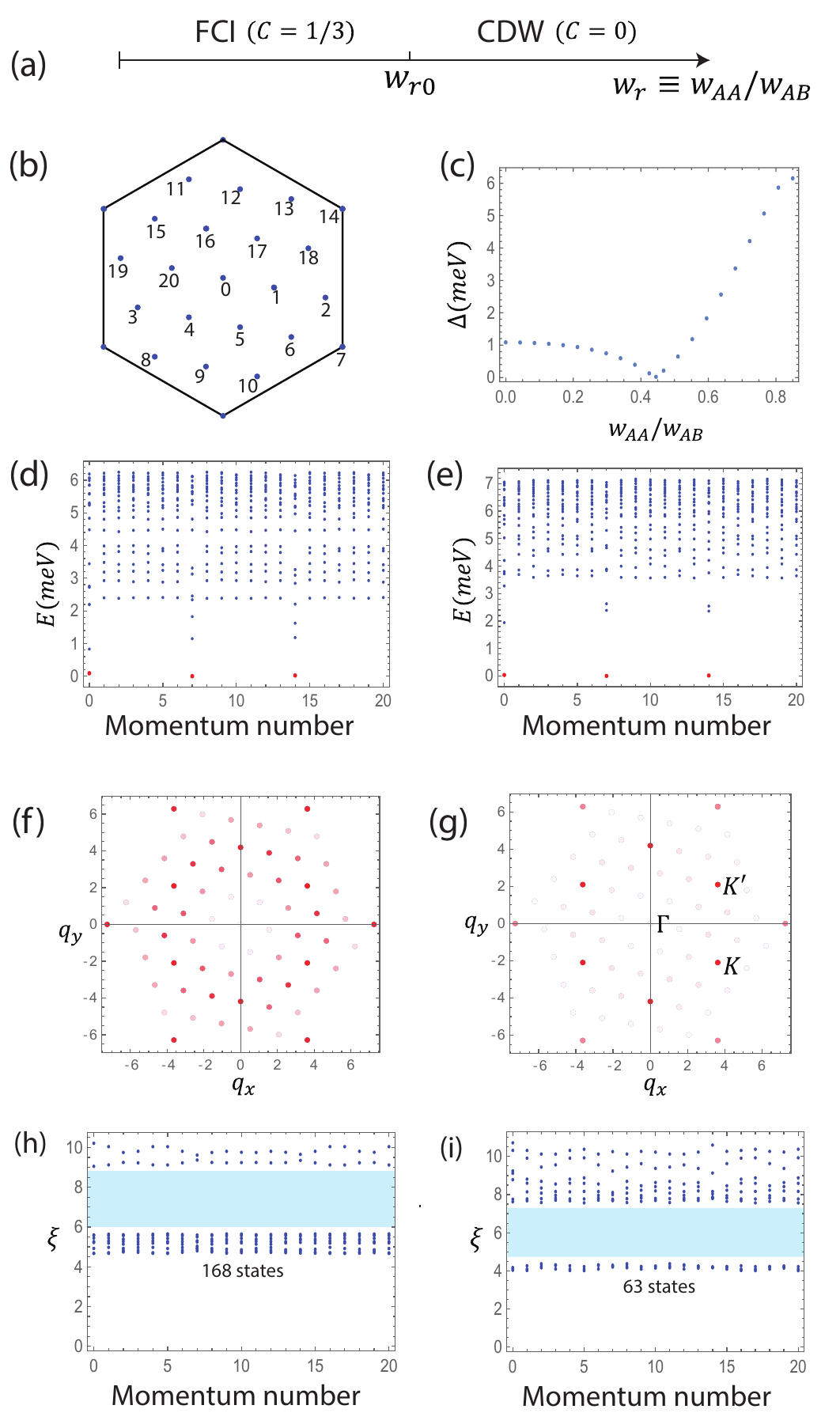}
\caption{(a) Phase diagram of TBG/hBN at different ratio $w_r\equiv w_{AA}/w_{AB}$. The chiral limit has $w_r=0$. At small $w_r$, the ground state is FCI with many-body Chern number $C=1/3$. At large $w_r>w_{r0}$, the ground state becomes CDW with $C=0$. The transition occurs at $w_{r0}\sim 0.45$. (b) Momentum points in the Brillouin zone of lattice 21t. (c) Energy gap $\Delta\equiv E_4-E_3$ of ED spectrum as a function of $w_r$. $E_{1\sim 3}$ are the energies of the three ground states, and $E_4$ is the lowest state above the ground states. (d), (f), (h) Energy spectrum, charge correlation function, and entanglement spectrum obtained from ED in lattice 21t with $w_r=0.3<w_{r0}$. The corresponding figures with $w_r=0.6>w_{r0}$ are shown in (e), (g), (i) respectively. Both the FCI and CDW phases have three degenerate ground states separated from higher states by a gap. Near $w_r\sim w_{r0}$ the gap closes and the phase transition occurs. The charge correlation function of CDW is highly peaked at $K$ points, in contrast to the FCI phase. The entanglement spectrum is calculated by separating the particles into $N_A=2$ and $N_B=5$ particles. The blue region denotes the gap in entanglement spectrum. The FCI phase should have 168 states below the gap.  }
\label{fig_FCICDW}
\end{figure}

{The results for filling at $\nu=1/3$ (one-third filling of one moir\'e band) are summarized in Fig. \ref{fig_FCICDW}. The ED computations shown here are performed in a lattice geometry denoted by 21t, which contains 21 momentum points as shown in Fig. \ref{fig_FCICDW}(b). The real space translational symmetries in lattice 21t are given by $\mathbf T_1=5\mathbf L_1-\mathbf L_2,\mathbf T_2=\mathbf L_1+4\mathbf L_2$, where $\mathbf L_1$ and $\mathbf L_2$ are moir\'e lattice vectors with an angle $60^\circ$ between them. This lattice geometry contains high symmetry $K$, $K'$ points, which are essential for distinguishing CDW and FCI.  When $w_{AA}/w_{AB}$ increases, the system transits from FCI at a small $w_{AA}$ to CDW at a large $w_{AA}$, which can be understood from the single particle band properties in Fig. \ref{fig_chiral}(b, c, d). We label the energies of the many-body states in ascending order $E_1\le E_2\le E_3\le \cdots \le E_n$. The FCI gap $\Delta\equiv E_4-E_3$ decreases with $w_{AA}/w_{AB}$ and vanishes at the transition point, as shown in Fig. \ref{fig_FCICDW}(c), and then the CDW gap increases with $w_{AA}/w_{AB}$.} Both the FCI and CDW at $v=1/3$ have three degenerate ground states. They can be distinguished using the density-density correlation, where the FCI is more or less uniform, while the CDW shows sharp peaks at $K$ and $K'$ in the Brillouin zone. The entanglement spectrum for the FCI state has a gap, and the number of states below the gap is consistent with the counting for the FCI state. The realistic value of $w_{AA}/w_{AB}$ for TBG is close to 0.7~\cite{Fuprx,PhysRevResearch.1.033126}, and thus the CDW is stabilized at zero magnetic fields. The FCI to CDW transition driven by $w_{AA}$ has been reported in Ref.~\onlinecite{Wilhelm2021f} based on exact diagonalization. Here we provide an understanding based on the emergent gauge field near the AA region.

\subsection{Role of SOC and external magnetic field}
One obvious difference between TBG and TMD is the SOC. In TMD, the Ising SOC locks the spin and valley, while in TBG, the SOC is negligible. Here, we introduce the SOC in TBG by proximating TBG to a substrate with strong SOC, such as $\mathrm{WSe_2}$. The proximity-induced SOC in TBG has been studied experimentally in the context of superconductivity, where SOC is found to be beneficial for stabilizing superconductivity~\cite{Lin2022sod}. An immediate question is whether SOC can also help stabilize the zero-field FCI. 

\begin{figure}[t]
\centering
\includegraphics[width=3.4 in]{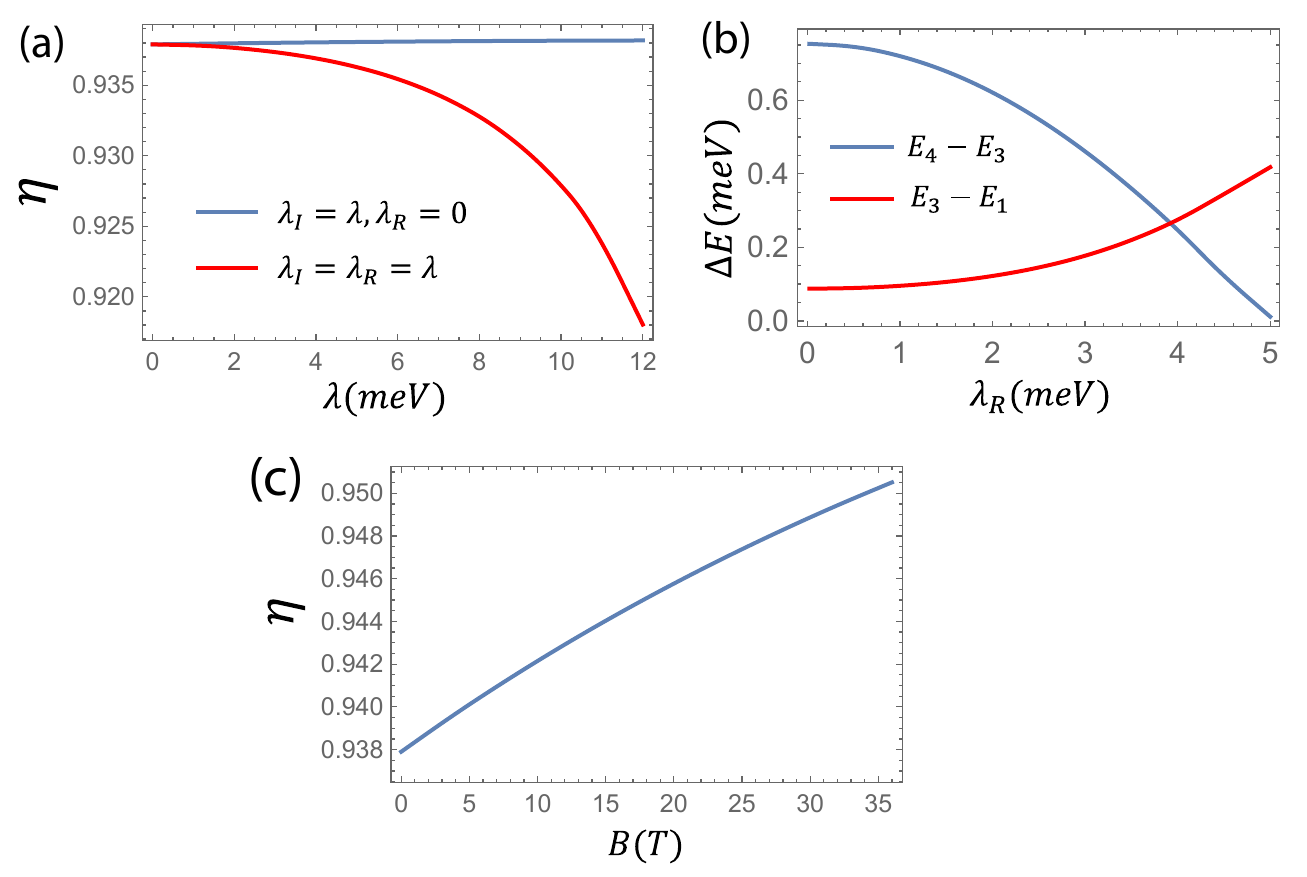}
\caption{(a) Overlap $\eta$ between eigenstates and the ideal zeroth Landau level as a function of Rashba SOC $\lambda_R$ and Ising SOC $\lambda_I$. (b) Many-body gap $E_4-E_3$ above the FCI ground states and energy separation between FCI ground states $E_3-E_1$ as a function of Rashba SOC computed from ED for TBG/hBN moir\'e lattice with $\lambda_I$ fixed at $7$ meV. (c) $\eta$ as a function of external magnetic field $B$. $\eta$ increases with external magnetic field, which makes the low energy states of TBG closer to Landau level and is beneficial for FCI. We take $w_{AA}=0.3w_{AB}$ all these plots.  }
\label{fig_etalmB}
\end{figure}

{We consider the Ising SOC $\lambda_I$ and Rashba SOC $\lambda_R$ contributions to the TBG/hBN system~\cite{PhysRevB.99.075438}. The Kane-Mele SOC is weak compared to other SOCs and is neglected. The effect of SOC induced by proximity to substrate is modeled by adding the following term to the graphene layer next to the substrate with strong SOC:
\be
h_\mathrm{SOC}=\lambda_I \zeta_z s_z+\lambda_R(\zeta_z\sigma_xs_y - \sigma_y s_x).
\ee
Here the Pauli matrices $\zeta$, $\sigma$, $s$ act on the valley, sublattice, and physical spin, respectively. The effect of hBN is given by a term $m\sigma_z$ added to the other graphene layer next to the aligned hBN. Because SOC is added to only one graphene layer, in the valley space $h_\mathrm{SOC}$ contains $\tau_0+\tau_z$ according to Eq. \eqref{eq3}. After the rotation of layer pseudospin in Eq. \eqref{eq4}, the $\tau_z$ term in $h_\mathrm{SOC}$ becomes $\tau_x$ which couples the $+\mathbf{a}$ and $-\mathbf{a}$ gauge fields. Therefore, the difference between the SOC strength in the two graphene layers leads to a coupling between opposite gauge fields. 

We compute the overlap $\eta$ with the ideal Landau level as a function of SOC strength in Fig. \ref{fig_etalmB}(a). It shows that $\eta$ decreases with the Rashba SOC, whereas $\eta$ is less sensitive to Ising SOC. This is because the zeroth Landau level stabilized by the $+\mathbf{a}$ and $-\mathbf{a}$  gauge fields are located at different sublattices, which can be coupled by the Rashba SOC. Therefore, the Rashba SOC leads to a direct coupling between the zeroth Landau levels formed by $\pm \mathbf{a}$ gauge fields, which increases the mixture of Landau levels and decreases $\eta$. On the contrary, the Ising SOC term cannot couple different sublattices, hence it can only induce Landau level mixing through higher Landau levels and its effect on $\eta$ is weaker. 

The effect of SOC is further supported by the spectrum of TBG/hBN moir\'e lattice computed via ED. At a small ratio $w_{AA}/w_{AB}=0.3<w_{r0}$ and with a finite SOC added to one graphene layer, there are three FCI ground states $E_1,E_2,E_3$ separated from higher states by a gap. We plot the gap $E_4-E_3$ and the energy difference between the FCI states $E_3-E_1$ as a function of $\lambda_R$ at a fixed $\lambda_I=7$ meV in Fig. \ref{fig_etalmB}(b). It shows that as $\lambda_R$ increases, the FCI gap decreases and separation between FCI states increases, indicating a gradual suppression of FCI by $\lambda_R$. This is consistent with Fig.\ref{fig_etalmB}(a) in which $\lambda_R$ directly couples the zeroth Landau levels stabilized by the opposite gauge fields.

We emphasize that the mixing of the zeroth Landau level by SOC is orginated from the difference between the SOC strengths in the two graphene layers. If the SOC terms are added identically to the two graphene layers, neither Rashba nor Ising SOC can lead to a significant change in $\eta$. }

We then study the role of the magnetic field $\mathbf{B}=\nabla\times\mathbf{A}$ in the absence of SOC. Applying an external magnetic field makes $\mathbf{k}\rightarrow \mathbf{k}- \mathbf{A}$ in Eq. \eqref{eq4}, which compensates for the opposite emerging gauge fields, thus favors the FCI state. The results of the dependence of $\eta$ on $B$ at zero SOC are shown in Fig. \ref{fig_etalmB}(c). It shows that $\eta$ increases with the external magnetic field, which is beneficial for FCI. This is consistent with the experimental observation that a magnetic field is required for FCI in TBG~\cite{Xie_Pierce2021}.

We remark that we only focus on FCI at zero temperature when considering the role of SOC. At finite temperature, the SOC may play an important role in stabilizing the FCI by providing the spin anisotropy in order to evade the Mermin-Wigner theorem. However, the effect of thermal fluctuations is beyond the ED calculations.

\section{Twisted TMD homobilayer}
Here we first review the low energy effective Hamiltonian for the twisted $\mathrm{MoTe_2}$ homobilayer starting from the AA aligned bilayer. The low-energy states contributing to the moir\'e bands are those states near the $K$ and $-K$ valley in the $\mathrm{MoTe_2}$ layer, where the $d_{x^2-y^2}\pm i\zeta d_{xy}$ and $d_{z^2}$ orbitals form a Dirac spectrum with a band gap of the order of 1.5 eV~\cite{PhysRevLett.108.196802}. In experiment, the moir\'e bands responsible for FCI are due to hybridization of the vallence band of the Dirac fermion in the presence of a moir\'e potential. Due to the large Dirac mass gap, the Dirac structure of the monolayer $\mathrm{MoTe_2}$ can be replaced by a simple hole band with parabolic dispersion, where the subtle Dirac nature of the low-energy states is neglected~\cite{PhysRevResearch.4.L032024}.  

The form of the Hamiltonian is governed by the $D_3$ point group symmetry generated by the three-fold rotation along the $z$ axis and a two-fold rotation along the $y$ axis that interchanges two layers. For the $\zeta=+$ valley, the Hamiltonian is \cite{PhysRevLett.122.086402}
\begin{align}\label{eq11n}
    \mathcal{H}_{+}= \begin{bmatrix}
-\frac{(\mathbf{k}-\mathbf{\kappa}_t)^2}{2m^*}+\Delta_b(\mathbf{r})  & \Delta_T(\mathbf{r})\\
\Delta_T(\mathbf{r})^\dagger &  -\frac{(\mathbf{k}-\mathbf{\kappa}_b)^2}{2m^*}+\Delta_t(\mathbf{r})
\end{bmatrix},
\end{align}
where $m^*=0.6m_e$ with $m_e$ being the bare electron mass. The top and bottom layer moir\'e potential has the form
\begin{align}
    \Delta_{b/t}(\mathbf{r})=2 V \sum_{j=1, 3, 5}\cos(\mathbf{G}_j\cdot \mathbf{r}\pm \psi),
\end{align}
and the interlayer tunneling
\begin{align}
    \Delta_{T}(\mathbf{r})=\omega (1+ e^{i \mathbf{G}_2\cdot \mathbf{r}}+  e^{i \mathbf{G}_3\cdot \mathbf{r}}),
\end{align}
where $\mathbf{G}_j=\frac{4\pi}{\sqrt{3}a_M} (\cos(j\pi/3),\ \sin(j\pi/3))$ with $a_M$ the moir\'e lattice constant. The Hamiltonian at $\zeta=-$ valley can be obtained by time-reversal operation. We can eliminate the momentum shift $\kappa_{t/b}$ in Eq. \eqref{eq11n} by a gauge transformation, after which $\Delta_T(\mathbf{r})$ is replaced by
\begin{align}
    \Delta_{T}(\mathbf{r})=\omega (e^{i \mathbf{q}_1\cdot \mathbf{r}}+ e^{i \mathbf{q}_2\cdot \mathbf{r}}+  e^{i \mathbf{q}_3\cdot \mathbf{r}}), 
\end{align}
while $\Delta_{b/t}(\mathbf{r})$ remains the same. Here $\mathbf{q}_1=\frac{4\pi}{\sqrt{3}a_M} (0,\ -1/\sqrt{3})$, $\mathbf{q}_j=\mathbf{G}_j+\mathbf{q}_1$. In terms of layer pseudospin $\tau$, the continuum model for the twisted TMD homobilayer becomes
\begin{equation}\label{eq7}
    \mathcal{H}(\mathbf{k})_+=-\frac{k^2}{2m^*}+d_0+\mathbf{d}(\mathbf{r})\cdot \mathbf{\tau},
\end{equation}
with texture vector given by the moir\'e potentials
\begin{align}
    \mathbf{d}(\mathbf{r})=\left(\mathrm{Re}\Delta_T^\dagger,\ \mathrm{Im} \Delta_T^\dagger,\ \frac{\Delta_b-\Delta_t}{2}\right),\ \ \ d_0=\frac{\Delta_b+\Delta_t}{2}.
\label{def_d}
\end{align}
It is known from density fucntional theory calculations that $\mathbf{d}(\mathbf{r})$ forms a skyrmion texture, which can be parameterized by two sphereical angles $\theta(\mathbf{r})$ and $\varphi(\mathbf{r})$, e.g. $\mathbf{d}(\mathbf{r})=|\mathbf{d}(\mathbf{r})|(\cos\theta\cos\varphi,\ \cos\theta\sin\varphi,\ \sin\theta)$. Eq. \eqref{eq7} is the standard double exchange model, which describes the coupling between the itinerant electrons/holes and the classical localized moments. The skyrmion texture can yield an emergent magnetic field for electrons~\cite{PhysRevLett.83.3737}. We perform a rotation in $\mathbf{\tau}$ to align $\mathbf{\tau}$ with the $\mathbf{d}$ vector.
\begin{equation}\label{eq11U}
    U(\mathbf{r})=\exp(-i \varphi(\mathbf{r})\tau_z/2)\exp(-i \theta(\mathbf{r})\tau_y/2),
\end{equation}
which generates an SU(2) gauge field $-i\partial_\mu\rightarrow -i\partial_\mu- a_\mu$ with $a_\mu\equiv i U^\dagger\partial_\mu U$. The Hamiltonian now has the form
\begin{equation}\label{eq12H}
    \mathcal{H}(\mathbf{k})_+=\frac{(-i \partial_\mu -a_\mu)^2}{2m^*}+d_0(\mathbf{r})+|\mathbf d(\mathbf{r})|\tau_z.
\end{equation}
In the large $|\mathbf{d}|$ limit, we may neglect the off-diagonal in $a_\mu$
\begin{equation}
    a_\mu \approx \partial_\mu \varphi \cos\theta \tau_z.
\end{equation}
In a unit cell, the total emergent gauge flux is $\int_{uc}dr^2 (\nabla\times \mathbf{a})_z=2\pi\tau_z$. For the moir\'e period in the experiment, the average emergent magnetic field strength is about 200 T. The problem is reduced to electrons moving in a strong but nonuniform magnetic field with a total flux $2\pi$ per unit cell. In addition, there is a nonuniform potential $d_0(\mathbf{r})$ and $d(\mathbf{r})$.  Therefore, FCI is generally expected when $d_0(\mathbf{r})$, $d(\mathbf{r})$ and $\nabla\times \mathbf{a}$ are sufficiently smooth, and $|d(\mathbf{r})|$ is large to avoid Landau level mixing.

In comparison to TBG, we also have two sets of Landau levels with opposite fields in TMD. We are interested in the topmost Landau level (the ``lowest" Landau level for hole). The difference between TMD and TBG is that now we have an exchagne field, $d(\mathbf{r})$ that shifts one set of Landau levels with respect to the other. This minimizes the hybridization of the topmost Landau levels with an opposite emergent magnetic field, and is beneficial for FCI. This implies that a stronger moir\'e potential, which can be achieved by applying pressure, is helpful for FCI. Explicit calculations in Ref.~\onlinecite{PhysRevResearch.5.L032022} indeed demonstrate that the pressure can help stabilize the FCI.

\subsection{Numerical results}
{We perform an ED calculation to explicitly verify the role of the moir\'e potential strength. The moir\'e potential strength can be characterized by a dimensionless ratio $\beta\equiv \bar d/E_K$ between the average value of $|\mathbf d(\mathbf r)|$ (denoted as $\bar d$) and a kinetic energy scale $E_{K}=\frac{\hbar^2 |G|^2}{2m^*}$, where $G$ is the moir\'e reciprocal lattice vector. We use the parameters from Ref. \onlinecite{Wang2023c} with $\theta=3.89^\circ$, $V=20.8$ meV, $\omega=-23.8$ meV, $\psi=107.7^\circ$, which gives $\beta=0.76$. We show that FCI is favored when $\beta$ is larger than some threshold value $\beta_c$, and this condition $\beta>\beta_c$ is indeed satisfied in realistic twisted MoTe$_2$ homobilayer, hence FCI can be realized in this material. In comparison, the FCI ground state in TBG/hBN has a similar requirement $w_r<w_{r0}$, which is not satisfied in realistic material, hence the FCI phase is not favored there.}

\begin{figure}[t]
\centering
\includegraphics[width=3.4 in]{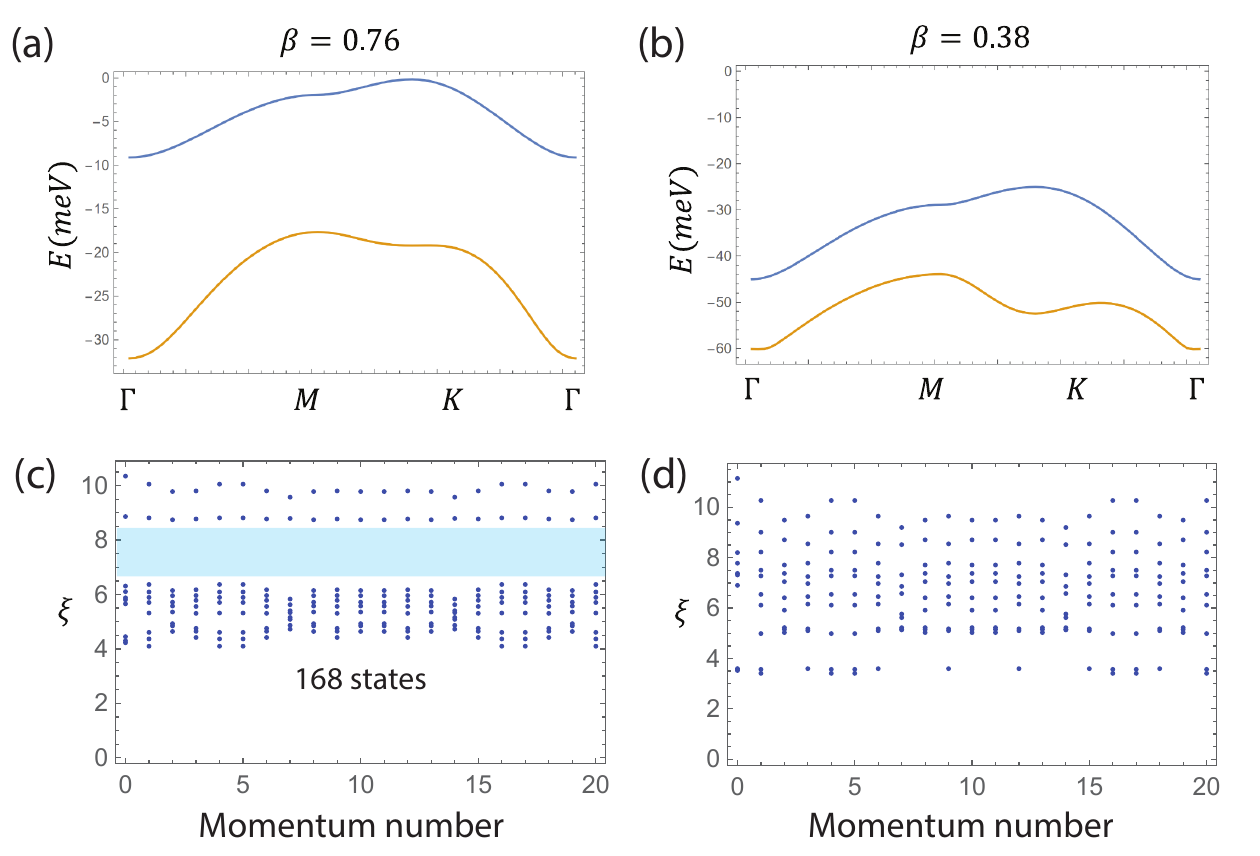}
\caption{(a), (b) Moir\'e band structure of twisted bilayer MoTe$_2$ with scaling coefficient $\beta=0.76$ and $\beta=0.38$ respectively, where $\beta\equiv \bar d/E_K$. $\beta=0.76$ corresponds to realistic systems. The top moir\'e band is used for ED computation. (c), (d) Entanglement spectrum of the three lowest states of twisted bilayer MoTe$_2$ with $\beta=0.76$ and $\beta=0.38$ respectively in lattice 21t. When $\beta=0.76$ there are 168 states below the entanglement gap denoted by the blue region ($N_A=2,N_B=5$), which features an FCI phase. When $\beta=0.38$, the entanglement gap closes. }
\label{fig_TMDES}
\end{figure}

{To demonstrate the importance of $\beta$, we compare the spectrum of systems with different values of $\beta$ by hypothetically scaling down the moir\'e potential $\mathbf d$ and $d_0$ by a constant. The single particle band structure for $\beta=0.76$ (realistic value) and $\beta=0.38$ is shown in Fig. \ref{fig_TMDES} (a,b) respectively, where a larger band gap exists for a larger $\beta$. Here, we focus on the Landau level mixing at the single particle level; therefore, we project the Coulomb interaction to the topmost band (blue band). The interaction has the same form as Eq. \eqref{eqHint} for TBG, with the density operator being replaced by $\rho_{\mathbf q}=\sum_{\mathbf k,\zeta}\lambda_{\zeta}(\mathbf k,\mathbf k+\mathbf q) \left(d_{\mathbf k, \zeta}^{\dagger} d_{\mathbf k+\mathbf q,\zeta}  \right)$, and the form factor is $\lambda_{\zeta}(\mathbf{k}, \mathbf{k}+\mathbf{q}) \equiv\left\langle u_{\mathbf k, \zeta} \mid u_{\mathbf k+\mathbf q, \zeta}\right\rangle$. We then diagonalize the interacting Hamiltonian numerically at the filling $\nu=1/3$. We find that the ground state is fully valley polarized. In the valley polarized sector, we find three ground states both for $\beta=0.76$ and $\beta=0.38$. We also compute the entanglement spectrum in Fig. \ref{fig_TMDES} (c,d). For $\beta=0.76$, the state count below the entanglement spectrum is consistent with the FCI phase; whereas for $\beta=0.38$, the entanglement gap is not well developed. We find that the closing of entanglement gap occurs at $\beta_c\sim0.55$. Therefore, the ground state at realistic parameter $\beta=0.76>\beta_c$ is FCI, whereas at smaller $\beta<\beta_c$ the FCI ground state breaks down. }

\begin{figure}[t]
\centering
\includegraphics[width=3.4 in]{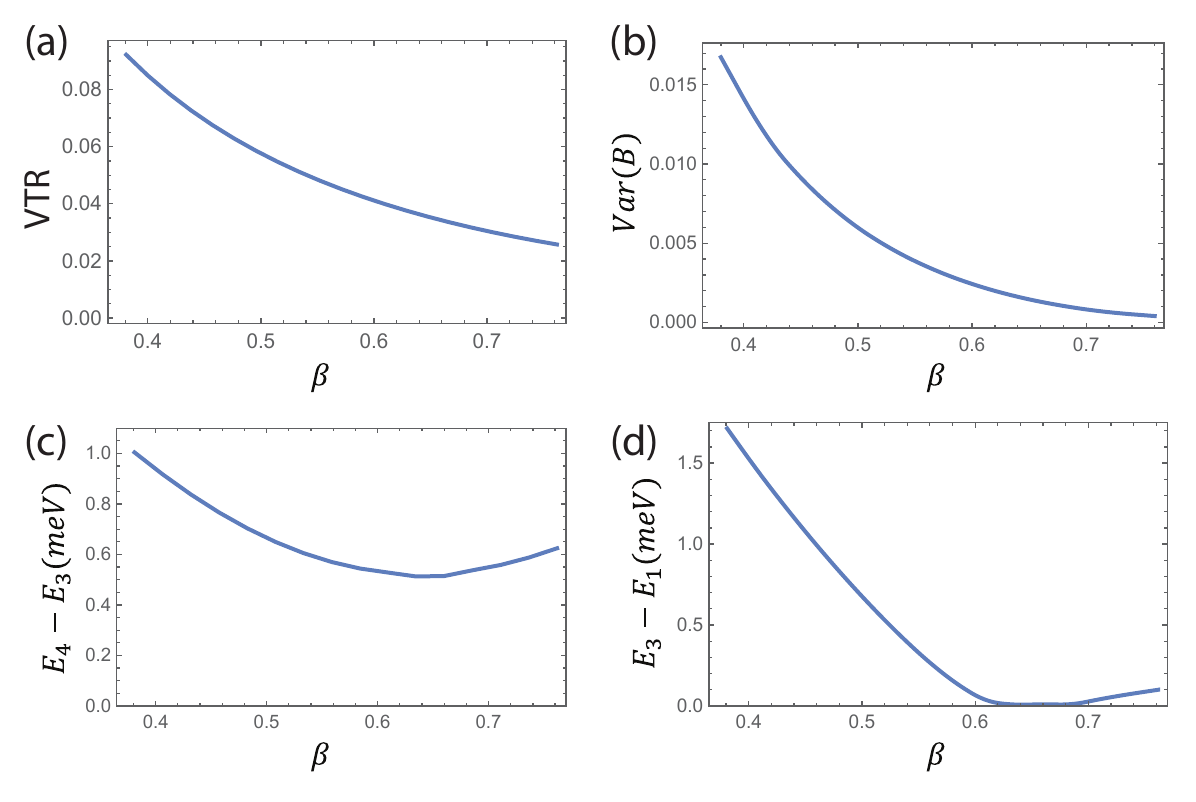}
\caption{(a) Violation of the trace condition and (b) variation of Berry curvature for the top moir\'e band of twisted bilayer MoTe$_2$ as a function of $\beta=\bar d/E_K$. (c) Energy difference between the fourth and third lowest energy manybody states as a function of $\beta$ obtained from ED. For FCI phase it characterizes the gap between the ground states and excited states. (d) Energy separation between the three lowest energy manybody states obtained from ED. As $\beta$ decreases, the violation of trace condition increases, and the system is no longer an FCI as indicated by the increased splitting of the first three states.  }
\label{fig_TMDscale}
\end{figure}

We further characterize the single particle band and the many-body ground state as a function of $\beta$ and the results are displayed in Fig. \ref{fig_TMDscale}. Both the variation in Berry curvature and the violation of the trace condition decrease with $\beta$, which can be understood in terms of the suppression of the Landau level mixing due to the exchange split. In the region where the FCI is fully developed $\beta>0.7$, the FCI gap increases with $\beta$ while the three ground states remain nearly degenerate, indicating a robust FCI.

\section{Discussion} 
In this work, we provide an intuitive picture of why the TBG does not support zero-field FCI based on the emergent U(1) gauge field picture. In TBG, near the AA region, the system can be mapped to two Dirac fermions coupled to an opposite emergent uniform magnetic field. The $w_{AA}$ interlayer tunneling causes hybridization between these two Dirac fermions. Although the hybridized band remains topological, the band geometry is not ideal for FCI due to the opposite emergent field. As a result, the FCI phase will be suppressed when $w_{AA}$ is strong. The effect of  $w_{AA}$ is perturbative in the sense that the hybridization of the two Dirac fermions is controlled by the ratio $w_{AA}/\hbar\omega_c$, where $\omega_c$ is the separation of the zeroth and the first Landau level. This is due to the sublattice polarization in the zeroth Landau level, which renders the direct hybridization between the zeroth Landau level vanishing. Thanks to the perturbative nature of $w_{AA}$, the FCI remains stable up to a critical value. The system transits into CDW when $w_{AA}$ is above this critical value. Unfortunately, $w_{AA}$ for realistic TBG is above the critical value and the zero-field FCI is absent. Applying a magnetic field can compensate for the opposite emergent gauge field and stabilize the FCI, consistent with experiments.

We study the role of the Rashba and Ising SOC in stabilizing the FCI in TBG, which can be introduced through the substrate. We find that the Ising SOC is perturbative due to the sublattice polarization of the zeroth Landau level. The effect of the Rashba SOC depends on how it is coupled to layers. When the Rashba SOC is introduced symmetrically for both TBG layers, it is also perturbative similar to that of the Ising SOC. However, when the Rashba SOC is introduced only in one layer, it suppresses the FCI due to the direct mixing of the zeroth Landau levels formed by the opposite gauge fields. 

We can also compensate for the gauge field $\mathbf{a}$ by introducing another emergent gauge field through the spin texture, analogously to the layer pseudospin skyrmion in the twisted TMD. Here we assume that the spin texture $\mathbf{S}(\mathbf{r})$ couples equally to the top and bottom graphene layer. This condition can be relaxed without changing the conclusion below. Then the diagonal elements of $\mathcal{H}_+$ in Eq. \eqref{eq4} becomes (neglecting SOC and sublattice potential for now)
\begin{equation}
     h_{t/b}(\mathbf{a})=-v_F(\mathbf{k}\pm\mathbf{a})\cdot\mathbf{\sigma} -J\mathbf{s}(\mathbf{r})\cdot \mathbf{S}(\mathbf{r}).
\end{equation}
We rotate the spin quantization axis to align $\mathbf{s}(\mathbf{r})$ with $\mathbf{S}(\mathbf{r})$ [see Eqs. \eqref{eq11U}, \eqref{eq12H}], which results in an emergent gauge field $b_{s,\mu}\equiv i U^\dagger\partial_\mu U$ that is the same for both layers
\begin{equation}
     U^\dagger h_{t/b}(\mathbf{a})U=-v_F(\mathbf{k}\pm\mathbf{a}-\mathbf{b}_s)\cdot\sigma - Js_z|\mathbf S(\mathbf{r})|.
\end{equation}
Due to $\mathbf{S}(\mathbf{r})$, the spin degeneracy of the band is lifted and we can partially fill the lower spin band. The curvature of $\mathbf{b}_s$ is just the skyrmion topological charge of $\mathbf{S}(\mathbf{r})$. In this way, we can stabilize FCI at zero magnetic field in TBG.

The discussion in terms of the U(1) gauge field is only valid near the AA region. However, this is sufficient to glean the nature of the topological flat band, because most states are localized near the AA region.  Recently, many more flat bands in the chiral limit have been realized in different systems~\cite{PhysRevLett.130.216401,Eugenio2023fs,Becker_Humbert_Zworski_2023,Le_Zhang_Fan_Wu_Chiu_2023,Sarkar2023c}. These models may not be recast into a form involving U(1) gauge field. Nevertheless, the same as the TBG, the hybridization between different chiral sectors suppresses the FCI in favor of CDW. For the realization of FCI, it is also desirable to minimize the coupling between the opposite chiral sectors.
In contrast, in TMD, the system can be regarded as electrons coupled to the layer pseudospin skyrmion texture through local exchange coupling. This yields an emergent U(1) gauge field coupled to the conduction electron, with the field strength given by one flux per moir\'e unit cell. (200 Tesla for twisted $\mathrm{MoTe_2}$ at an angle $3.7^\circ$). The local exchange coupling splits the originally degenerate bands into two bands with opposite U(1) gauge fields. Furthermore, the two bands are separated by an energy gap given by the local exchange coupling. When the local exchange coupling is stronger than the fermi energy, the hybridization between the bands with opposite gauge fields is suppressed. Therefore, the lower band closely resembles the Landau level if the emergent gauge field is sufficiently uniform in space, which sets the stage for the emergence of zero field FCI as observed in the experiment.

On the basis of this understanding, we discuss strategies to realize topological flat bands suitable for FCI. In condensed matter systems, there are many examples of conduction electrons coupled to an emergent gauge field due to the texture of the order parameter. One prime example is the case with magnetic skyrmion. In this context, it is required to optimize the texture in order to obtain a uniform emergent magnetic field. It is also better when the local coupling between the conduction electron and the texture is large, such that the bands with opposite Chern numbers are well separated in energy to minimize the coupling between these bands. Another direction is that we may use the Dirac fermion coupled to the emergent gauge field. Generally, the emergent magnetic field is nonuniform. However, this nonuniform magnetic field still gives an exactly flat zeroth Landau level for the Dirac fermion \cite{Paul_Zhang_Fu_2023}, which is helpful for the realization of FCI. In a model like TBG where the chiral limit exists, one needs to minimize the coupling between different chiral sectors. This can be achieved either by reducing the Dirac fermion velocity to enhance the strength of the emergent magnetic field or by reducing $w_{AA}$, because the hybridization depends on $w_{AA}/\hbar\omega_c$. We would also like to remark that the pentalayer graphene on hBN cannot fit into the current framework in terms of an emergent gauge field~\cite{Lu_Han_Ju_2023}. There are also models that host FCI with a periodic emergent gauge field but with zero average, as realized in graphene subjected to periodic strain \cite{PhysRevLett.131.096401,PhysRevLett.130.196201,PhysRevB.108.125129}, which cannot be captured by the current discussion.

Finally, we would like to mention a recent nice work that provides a unified description of TBG and TMD in terms of perturbed Dirac field, which can be rewritten as a Dirac fermion coupled with an emerging gauge field~\cite{Crepel_Regnault_Queiroz_2023}. In the chiral limit, the appearance of the flat band is a manifestation of the chiral anomaly and the related Atiyah-Singer index~\cite{Parhizkar_Galitski_2023}. It is shown that the deviation from the chiral limit in TMD is weaker than that for TBG due to significant corrugation of the TMD layers, and hence the topological flat band is more suitable for FCI. Ref. \onlinecite{Crepel_Regnault_Queiroz_2023} provides a different perspective on why no zero-field FCI occurs in TBG.

\begin{acknowledgements}
{\it Acknowledgments.---} The work at LANL was carried out under the auspices of the U.S. DOE NNSA under contract No. 89233218CNA000001 through the LDRD Program, and was performed, in part, at the Center for Integrated Nanotechnologies, an Office of Science User Facility operated for the U.S. DOE Office of Science, under user proposals $\#2018BU0010$ and $\#2018BU0083$. KS is supported by the Office of Naval Research Grant No. N00014-21-1-2770, 
Air Force Office of Scientific Research MURI FA9550-23-1-0334 the Gordon and Betty Moore Foundation Grant No. GBMF10694. HL and HYK acknowledge the support from the Natural Sciences and Engineering Research Council of Canada Discovery Grant No. 2022-04601 and the Canada Research Chairs Program. YBK is supported by the NSERC of Canada and the Center for 
Quantum Materials at the University of Toronto. Computations were performed on the Niagara supercomputer at the SciNet HPC Consortium. SciNet is funded by: the Canada Foundation for Innovation under the auspices of Compute Canada; the Government of Ontario; Ontario Research Fund - Research Excellence; and the University of Toronto.

\end{acknowledgements}

\appendix

\setcounter{figure}{0}
\setcounter{table}{0}
\makeatletter
\renewcommand{\thefigure}{S\arabic{figure}}

\section{Conditions for the U(1) description in TBG}\label{AppenA}

Here, we discuss the conditions under which the TBG model can be mapped to Dirac fermion coupled to an emergent U(1) gauge field. The Hamiltonain at $+$ valley can be written as
\begin{equation}\label{Hk}
    \mathcal{H}(k)_+
{=-v_F (\mathbf{k} - \mathbf{a})\cdot \mathbf{\sigma} + \mathbf{d}\cdot\mathbf{\tau} \sigma_0},
\end{equation}
{where $\mathbf{a}=(\mathbf{a}_x\cdot\mathbf{\tau},\ \mathbf{a}_y \cdot \mathbf{\tau})$ is the emergent SU(2) gauge field due to the interlayer tunneling near the AB region with the nonzero components $a_x^x = w_{AB} \sum_{j=1}^3\cos(\mathbf{q}_j\cdot\mathbf{r})\cos(\mathbf{q}_j\cdot\mathbf{r}_{AB})/v_F$, $a_x^y = -w_{AB} \sum_{j=1}^3\sin(\mathbf{q}_j\cdot\mathbf{r})\cos(\mathbf{q}_j\cdot\mathbf{r}_{AB})/v_F$, $a_y^x = w_{AB} \sum_{j=1}^3\cos(\mathbf{q}_j\cdot\mathbf{r})\sin(\mathbf{q}_j\cdot\mathbf{r}_{AB})/v_F$, and $a_y^y = -w_{AB} \sum_{j=1}^3\sin(\mathbf{q}_j\cdot\mathbf{r})\sin(\mathbf{q}_j\cdot\mathbf{r}_{AB})/v_F$. Here the Pauli matrices $\tau$ act on the layer pseudospin. The last term is induced by the interlayer tunneling near the AA region and behaves as an exchange interaction between the layer pseudospin and a vector field $\mathbf{d}=(w_{AA} \sum_{j=1}^3\cos(\mathbf{q}_j\cdot\mathbf{r}),\  -w_{AA} \sum_{j=1}^3\sin(\mathbf{q}_j\cdot\mathbf{r}),0)$. To align the layer pseudospin with the vector field, we first perform a local gauge transformation such that 
\begin{equation}\label{UHU}
    U^\dagger H(\mathbf{k})_+U = - v_F\left( \mathbf{k} - iU^\dagger \nabla U - U^\dagger \mathbf{a} U \right) \cdot \mathbf{\sigma}
+ |\mathbf{d}|\tau_z \sigma_0,
\end{equation}
with the local layer pseudospin rotation operator $U=\exp\left( -i\frac{\theta}{2} \mathbf{n}_0 \cdot \tau\right)$. Here $\mathbf{n}_0=(-\sin\phi,\ \cos\phi,0)$ where $\theta$ and $\phi$ are the azimuthal and polar angle of the $\mathbf{d}$ vector. Note that $\mathbf{d}$ is always in-plane and $\theta=\pi/2$ is fixed. Now we have a new SU(2) gauge field,
\begin{equation}\label{SU2}
\begin{split}
    \mathbf{a}' &= iU^\dagger \nabla U + U^\dagger \mathbf{a} U \\
    & = - \frac{\nabla \phi }{2}
\begin{pmatrix}
1 &   e^{-i\phi} \\
e^{i\phi} & -1
\end{pmatrix}
+ 
\frac{1}{2}
\begin{pmatrix}
1 &   e^{-i\phi} \\
-e^{i\phi} & 1
\end{pmatrix} 
\mathbf{a}
\begin{pmatrix}
1 & -e^{-i\phi} \\
e^{i\phi} & 1
\end{pmatrix}. 
\end{split}
\end{equation}
Even though $\mathbf{d}(\mathbf{r})$ lies in the same plane in the $\mathbf{\tau}$ space, we still obtain nontrivial topology for the Dirac fermion due to the presence of $\mathbf{a}$. The situation is similar to the case with coplanar spin texture, where SOC is required to have a nonzero Berry curvature. 

\begin{figure}[t]
\centering
\includegraphics[width=3.4 in]{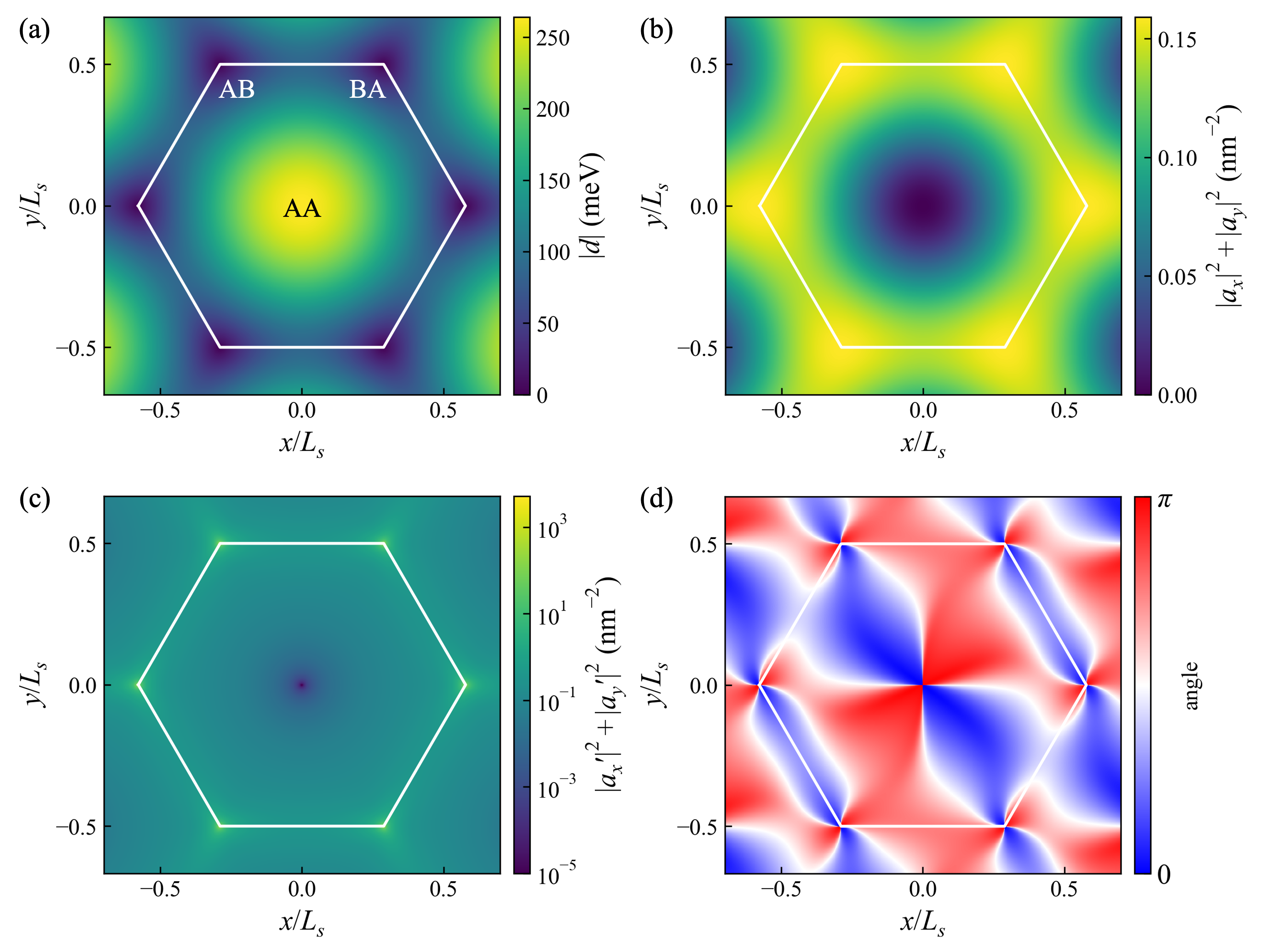}
\caption{{(a) and (b) Magnitudes of the emergent vector field $|\mathbf{d}|$ and the SU(2) gauge field $|\mathbf{a}_x|^2+|\mathbf{a}_y|^2$ in Eq. (\ref{Hk}), respectively. The white hexagon encloses the Moir{\'e} unit cell. The high-symmetry AA and AB/BA stacking points are marked in (a). (c) The magnitude of the SU(2) gauge field $|\mathbf{a}_x'|^2+|\mathbf{a}_y'|^2$ in Eq. (\ref{SU2}). (d) The angle between the two components $\mathbf{a}_x'$ and $\mathbf{a}_y'$ of the emergent SU(2) gauge field. Near the AA region the angle is either 0 or $\pi$, hence $\mathbf{a}_x'$ and $\mathbf{a}_y'$ are parallel and a U(1) description is possible.  }}
\label{gauge}
\end{figure}

To further simplify the problem, we consider the condition $\mathbf{a}_x'\parallel \mathbf{a}_y'$ in which the SU(2) gauge field can be reduced to U(1) where $\mathbf{a}'=(\mathbf{a}_x'\cdot\mathbf{\tau},\ \mathbf{a}_y'\cdot\mathbf{\tau})$. In particular, we find that this condition can only be satisfied around the AA region. To demonstrate this result, we plot the magnitudes of the vector field $|\mathbf{d}|$ and the SU(2) gauge field $|\mathbf{a}_x|^2+|\mathbf{a}_y|^2$, as shown in Figs. \ref{gauge}(a) and (b), respectively. Apparently, $|\mathbf{d}|$ equals zero at the AB/BA point where the last term in Eq. (\ref{Hk}) vanishes. On the other hand,  $|\mathbf{a}_x|^2+|\mathbf{a}_y|^2$ is zero at the AA stacking point where the SU(2) gauge field $\mathbf{a}$ in Eq. (\ref{Hk}) vanishes. Apart from these high-symmetry points, we have nonzero vector field $\mathbf{d}$ as well as the SU(2) gauge field $\mathbf{a}$. For the new gauge field $\mathbf{a}'$, we show its magnitude $|\mathbf{a}_x'|^2+|\mathbf{a}_y'|^2$ and the angle between two components $\mathbf{a}_x'$ and $\mathbf{a}_y'$ in Fig. \ref{gauge}(c) and (d), respectively. {The intermediate angle between 0 an $\pi$} indicates that the new SU(2) gauge field $\mathbf{a}'$ cannot be reduced to U(1) other than those high-symmetry points around which {the angle between $\mathbf{a}_x'$ and $\mathbf{a}_y'$ winds, as shown in Fig. \ref{gauge}(d).} 
To understand this behavior, we note that $\mathbf{d}$ vanishes at AB/BA point and hence its polar angle $\phi$ is ill-defined. Therefore, we don't need to perform the gauge transformation in Eq. (\ref{UHU}) at AB/BA where $\mathbf{a}_x \perp \mathbf{a}_y$, indicating that the SU(2) gauge field cannot be reduced to U(1). However, at the AA point, both $\mathbf{a}$ and $\mathbf{a}'$ vanish that yields the central vortex in Fig. \ref{gauge}(c). {Slightly away from the AA region, the angle between $\mathbf{a}_x'$ and $\mathbf{a}_y'$ is either $0$ or $\pi$, and an U(1) description is possible.}  }

\section{Landau level formulation of TBG}\label{AppenB}
\begin{figure}[b]
\centering
\includegraphics[width=3.4 in]{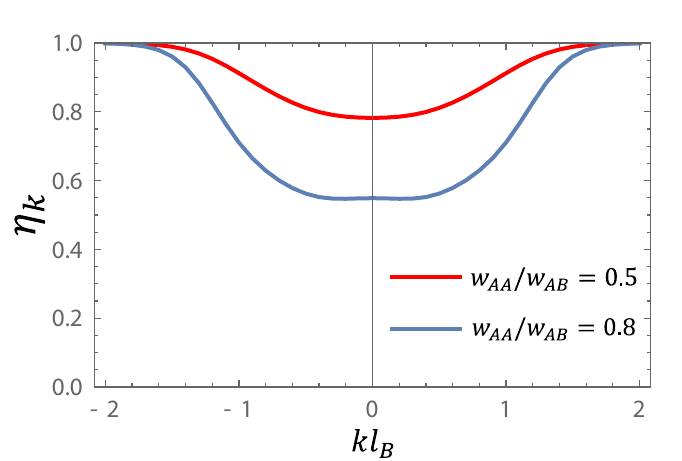}
\caption{Overlap $\eta_k$ between the eigenstate and ideal Landau level at different ratio $w_{AA}/w_{AB}$.  }
\label{fig_overlapLD}
\end{figure}

The Hamiltonian of TBG near the AA region can be regarded as two coupled Dirac fermions under opposite gauge fields as in Eq. \eqref{eq4}. The spectrum can be solved by introducing raising and lowering operators, similar to a Landau level problem. We choose the Landau gauge $\mathbf{a}=\frac{4\pi w_{AB}}{L_s ev_F}(y,\ 0),\ b=(\nabla \times \mathbf{a})_z\sim 120$ T, and let $l_B=\sqrt{\frac{\hbar v_F L_s}{4\pi w_{AB}}},\ \hbar\omega_c=\sqrt{\frac{8\pi\hbar v_F w_{AB}}{L_s}}$. We define the following operators for the upper and lower blocks with opposite gauge field, respectively:
\bea
a&=&\frac{1}{\sqrt{2}}(i\hat k_y l_B+\frac{y}{l_B}-\hat k_x l_B),\ a^\dagger=\frac{1}{\sqrt{2}}(-i\hat k_y l_B+\frac{y}{l_B}-\hat k_x l_B),  \nonumber\\
a'&=&\frac{1}{\sqrt{2}}(i\hat k_y l_B+\frac{y}{l_B}+\hat k_x l_B),\ a'^\dagger=\frac{1}{\sqrt{2}}(-i\hat k_y l_B+\frac{y}{l_B}+\hat k_x l_B). \nonumber\\
\eea
Then the upper and the lower blocks of Eq. \eqref{eq4} corresponding to $\pm b$ gauge fields can be rewritten as
\bea
H_{\mathrm{upper}}&=&
    \begin{bmatrix}
0  & -\hbar \omega_c a^\dagger \\
-\hbar \omega_c a &  0
\end{bmatrix}, \nonumber\\
H_{\mathrm{lower}}&=&
    \begin{bmatrix}
0 & \hbar \omega_c a' \\
\hbar \omega_c a'^\dagger &  0
\end{bmatrix}.
\eea
Within the Landau gauge, the momentum along the $x$ direction is a good quantum number, which is denoted as $k$. The states created by $a^\dagger$ and $a'^\dagger$ are given by:
\bea
\langle \mathbf r \ket{n,k,+}&=&\frac{1}{2^{n/2}\sqrt{n!}\pi^{1/4}}e^{ikx}H_n(\frac{y}{l_B}-l_Bk)e^{-\frac{1}{2}(\frac{y}{l_B}-kl_B)^2}, \nonumber\\
\langle \mathbf r \ket{n,k,-}&=&\frac{1}{2^{n/2}\sqrt{n!}\pi^{1/4}}e^{ikx}H_n(\frac{y}{l_B}+l_Bk)e^{-\frac{1}{2}(\frac{y}{l_B}+kl_B)^2}.
\eea
Here $H_n(x)$ is the Hermite polynomial. These states satisfy
\bea
\ket{n,k,+}&=&\frac{(a^\dagger)^n}{\sqrt{n!}}|0,k,+\rangle, \nonumber\\
\ket{n,k,-}&=&\frac{(a'^\dagger)^n}{\sqrt{n!}}|0,k,-\rangle.
\eea
To construct the eigenstates of Eq. \eqref{eq4}, we need to take into account spin and sublattice. For the upper block with $+b$ field, we define four states:
\bea
\phi_{n,k,+,1}&=&\begin{pmatrix}
\ket{n,k,+}  \\ 0\\0\\0
\end{pmatrix},\  \phi_{n,k,+,2}=\begin{pmatrix}
0\\\ket{n-1,k,+}  \\0\\0
\end{pmatrix}, \nonumber\\
\phi_{n,k,+,3}&=&\begin{pmatrix}
0\\0\\ \ket{n,k,+}  \\ 0
\end{pmatrix},\  \phi_{n,k,+,4}=\begin{pmatrix}
0  \\0\\0 \\ \ket{n-1,k,+}
\end{pmatrix},
\eea
where the four components of each state represent $(\uparrow A,\uparrow B, \downarrow A, \downarrow B)$ respectively. Note that when $n=0$ only $\phi_{n,k,+,1}$ and $\phi_{n,k,+,3}$ exist. Using the basis $\{\phi_{n,k,+,1},\phi_{n,k,+,2},\phi_{n,k,+,3},\phi_{n,k,+,4}  \}$, and taking into account the effect of SOC and the proximity of the hBN layer, the Hamiltonian for the $+b$ block becomes 
\bea
H(n,+)=\hbar \omega_c \begin{pmatrix}
1 & 0\\
0 & 1
\end{pmatrix}\otimes \begin{pmatrix}
0 & -\sqrt{n}\\
-\sqrt{n} & 0
\end{pmatrix}.
\eea
For the lower block with $-b$ field, similarly we define:
\bea
\phi_{n,k,-,1}&=&\begin{pmatrix}
\ket{n-1,k,-}  \\ 0\\0\\0
\end{pmatrix},\  \phi_{n,k,-,2}=\begin{pmatrix}
0\\\ket{n,k,-}  \\0\\0
\end{pmatrix}, \nonumber\\
\phi_{n,k,-,3}&=&\begin{pmatrix}
0\\0\\ \ket{n-1,k,-}  \\ 0
\end{pmatrix},\  \phi_{n,k,-,4}=\begin{pmatrix}
0  \\0\\0 \\ \ket{n,k,-}
\end{pmatrix},
\eea
and the Hamiltonian for the $-b$ block becomes
\bea
H(n,-)=\hbar \omega_c \begin{pmatrix}
1 & 0\\
0 & 1
\end{pmatrix}\otimes \begin{pmatrix}
0 & \sqrt{n}\\
\sqrt{n} & 0
\end{pmatrix}.
\eea
The $\pm b$ blocks are mixed by the $w_{AA}$ term. The mixing term has an integral of the form $w_{AA}\langle n, k, +\ket{n',k',-}$. The integral along the $x$ direction is nonzero only when $k=k'$. Hence the mixing term has a $k$ dependence of the form $\langle n, k, +\ket{n',k',-}\sim \delta_{k,k'} \int e^{-\frac{1}{2}(\frac{y}{l_B}-kl_B)^2}e^{-\frac{1}{2}(\frac{y}{l_B}+kl_B)^2}dy$ which decays rapidly at a large $k$.

The full Hamiltonian is built by taking into account the effect of SOC, the hBN layer and the mixing caused by $w_{AA}$. When $w_{AA}=0$, there are four low energy eigenstates at each $k$ coming from the zeroth Landau level of $\pm b$ blocks with both spin orientations. When $w_{AA}$ increases, each eigenstate is no longer fully within the $+b$ or $-b$ block, and is a mixture of the Landau levels at opposite gauge fields. We can measure the overlap between the eigenstate $\psi_k$ and the pure zeroth Landau level by:
\bea
\eta_k&=&max\{ \eta_{k,+}\ ,\ \eta_{k,-} \}, \nonumber\\ \eta_{k,\pm}&=&|\langle \psi_k|0,k,\pm,\uparrow\rangle  |^2+|\langle \psi_k|0,k,\pm,\downarrow\rangle  |^2.
\eea
We plot $\eta_k$ at $w_{AA}/w_{AB}=0.5,0.8$ as a function of $k$ in Fig.\ref{fig_overlapLD}. When $w_{AA}$ is small, $\eta_k$ is close to 1, and it deviates from 1 when $w_{AA}$ increases. To quantify the overlap between the eigenstate at finite $w_{AA}$ and the ideal Landau level, we define $\eta$ as the average of $\eta_k$ within $|k|<1/l_B$. This is the $\eta$ used in the main text. The dependence of $\eta$ on $w_{AA}$ demonstrates that $w_{AA}$ mixes Landau levels with opposite gauge fields and makes the eigenstates deviate from the ideal Landau levels, which is unfavorable for the FCI phase.


%

\end{document}